\documentclass[aps,prc,showpacs,preprint,numbers,floatfix,showkeys]{revtex4}
\usepackage{amsmath,amssymb,amsfonts}
\usepackage{graphicx}
\usepackage{color}
\allowdisplaybreaks
\begin{document}
\title{
Relativistic Expansion of Electron-Positron-Photon Plasma Droplets
and Photon Emission}
\author{R. \surname{Yaresko$^{1,2}$}} 
\email{r.yaresko@fzd.de}
\author{Munshi G. \surname{Mustafa$^{1,3}$}} 
\email{musnhigolam.mustafa@saha.ac.in}
\author{B. \surname{K\"ampfer}$^{1,2}$}
\email{kaempfer@fzd.de}
\affiliation{
$^1$Forschungszentrum Dresden-Rossendorf, 01314 Dresden, Germany \\ 
$^2$Institut f\"ur Theoretische Physik, TU~Dresden, 01062 Dresden, Germany\\
$^3$Theory Division, Saha Institute of Nuclear Physics, 1/AF Bidhannagar, Kolkata 700064, India}

\begin{abstract}
The expansion dynamics of hot electron-positron-photon plasma droplets
is dealt with within relativistic hydrodynamics. Such droplets, envisaged to be
created in future experiments
by irradiating thin foils with counter-propagating ultra-intense laser
beams, are sources of flashes of gamma radiation.
Warm electron-positron plasma droplets may be identified and characterized by
a broadened 511 keV line.
\end{abstract}
\pacs{12.20.-m, 52.50.Jm, 52.27.Ny, 52.25.Os}
\keywords{Quantum Electrodynamics, Laser, Relativistic plasma, Gamma radiation and absorption}
\maketitle

\section{Introduction}

The rapid progress in ultra-intense laser technology \cite{lasertechnology}
triggered the interest in properties of relativistic plasmas
consisting mainly of electrons ($e^-$), positrons $e^+$ and photons ($\gamma$). 
For instance, in \cite{Rafelski1,Munshi_bk,Rafelski2,Ruffini}
various processes, including thermal (kinetic) and chemical equilibration,
have been considered for a hot charge-symmetric electron-positron-photon 
($e^- e^+ \gamma$) plasma,
while the kinetics of a temporarily created
electron-positron plasma has been analyzed in \cite{Blaschke}. 
Hot plasmas, may be charge-symmetric or asymmetric 
or with ion load, have interesting features \cite{Thoma}, e.g.\
with respect to collective excitation modes (including the van Hove singularity),
astrophysical situations and secondary particle (such as muon or pion)
production. Furthermore, an optically thick $e^- e^+ \gamma$ plasma represents
the Abelian analog of the non-Abelian quark-gluon plasma \cite{QM} currently
investigated at large-scale accelerators.

The scenario for droplet creation can be visualized in the following way: 
Suppose that the radiation
pressure of counter-propagating ultra-intense laser beams on a thin foil
is so high that a substantial compression accompanied by sizeable
pair creation is achieved. Estimates for such a process have been presented 
in \cite{Shen_MtV}. Extrapolating these estimates one may envisage the
idealized scenario of a hot and optically thick $e^- e^+ \gamma$ plasma,
with temperatures exceeding the 1 MeV scale.  
Optically thick $e^- e^+ \gamma$ plasmas play an important role in the 
gamma ray bursts phenomena \cite{Ruffini2}.

In \cite{Munshi_bk} it has been pointed out that an $e^- e^+ \gamma$ plasma droplet
at initial temperatures $T$ much larger than the electron mass $m$ emits an intense
gamma ray burst. However, the expansion dynamics was considered in
\cite{Munshi_bk} in a schematic manner. Nevertheless, 
the time structure and the spectrum of the burst depend crucially on
the expansion dynamics of the plasma droplet driven by the thermodynamic
pressure gradients. 
It is necessary to consider a realistic expansion dynamics as well as the particle 
production rates in the relativistic framework for a proper understanding
of $e^- e^+ \gamma$ plasmas produced in the laboratory and astrophysical sites
such as gravitational collapse. 
The aim of the present paper is to quantify the droplet
expansion dynamics by means of the relativistic hydrodynamics. This provides 
a realistic evaluation of the time evolution of the temperature profile and
velocity profile. The latter one is important for properly accounting for
the blue/red shift effects of the emitted photons.

Another avenue towards the creation of $e^- e^+$ plasma droplets has been
envisaged in \cite{Blaschke} as a result of the dynamical Schwinger
process. Due to the short (sub cycle) time scales the thermalization
and the collective dynamics are not an issue. Instead, the emergence
of $e^- e^+$ annihilation into $\gamma$ pairs may be employed as diagnostic
tool.   

Our paper is organized as follows. In section \ref{sec2} we consider the spherically
symmetric expansion of plasma droplets within the relativistic hydrodynamics.
This input is used in section \ref{sec3} to estimate the spectrum of emitted photons.
Section \ref{sec4} gives an account of the spectral emissivity for a warm plasma with 
a temperature in the order of the electron mass and focusses on the onset 
of the broadening of the 511 keV line when achieving higher temperatures.
Section \ref{sec5} is devoted to the summary. 
Appendix \ref{appA} presents the expansion dynamics with planar geometry. 

\section{Relativistic Hydrodynamics\label{sec2}}

A sufficiently large $e^- e^+ \gamma$ plasma droplet can be dealt with by
means of relativistic hydrodynamics. The equations of motion imply local
energy and momentum conservation expressed as
\begin{equation}
{T^{\mu \nu}}_{; \mu} = 0,
\label{eq.1}
\end{equation}
where $\mu, \nu = 0, \dots, 3$ denote Lorentz indices and a colon stands
for the covariant derivative. For an isotropic medium (gas or fluid or plasma)
the energy-momentum tensor reads
\begin{equation}
T^{\mu \nu} = (e + p) u^\mu u^\nu - p g^{\mu \nu} + \cdots.
\label{eq.2}
\end{equation}
Here, the four-velocity is $u^\mu$ with normalization $u^\mu u_\mu = 1$
and $g^{\mu \nu}$ denotes the metric tensor,
${\rm diag}(g^{\mu \nu}) = (1, -1, -1, -1)$ in Cartesian coordinates.
The energy density is $e$, and $p$ stands for the thermodynamic pressure.
The ellipsis $\cdots$ in (\ref{eq.2}) denote dissipative terms being
proportional to transport coefficients, such as shear and bulk viscosities,
for instance. Neglecting the latter ones, (\ref{eq.1}, \ref{eq.2}) can be
solved once an equation of state is given. For $T > m$, the equation of state
of a relativistic ideal gas can be employed, that is $e = 3p$.
We consider a charge-symmetric plasma with zero net density. Accordingly,
the chemical potential is zero, too.
The pressure and temperature for our equation of state are related via
$p = d_{eff} \frac{\pi^2}{90} T^4$
(we employ units with $\hbar = c = 1$)
with $d_{eff} = 2 + \frac78 4$ for photons (first term) and electrons and positrons
in equilibrium. This is the leading-order term in an expansion in
powers of $m/T$.

Special symmetries simplify (\ref{eq.1}) tremendously. We consider here
spherically symmetric droplets (planar slab geometry is considered in appendix A).
Then (\ref{eq.1}) degenerates to two coupled equations, one for the pressure
gradient and one for the radial velocity gradient. These equations can
be solved by the methods of characteristics (cf.~\cite{Baym} for details). 
Numerical results are exhibited in Fig.~1.

\begin{figure}[!htbp]
\begin{center}
\includegraphics[width=0.49\columnwidth]{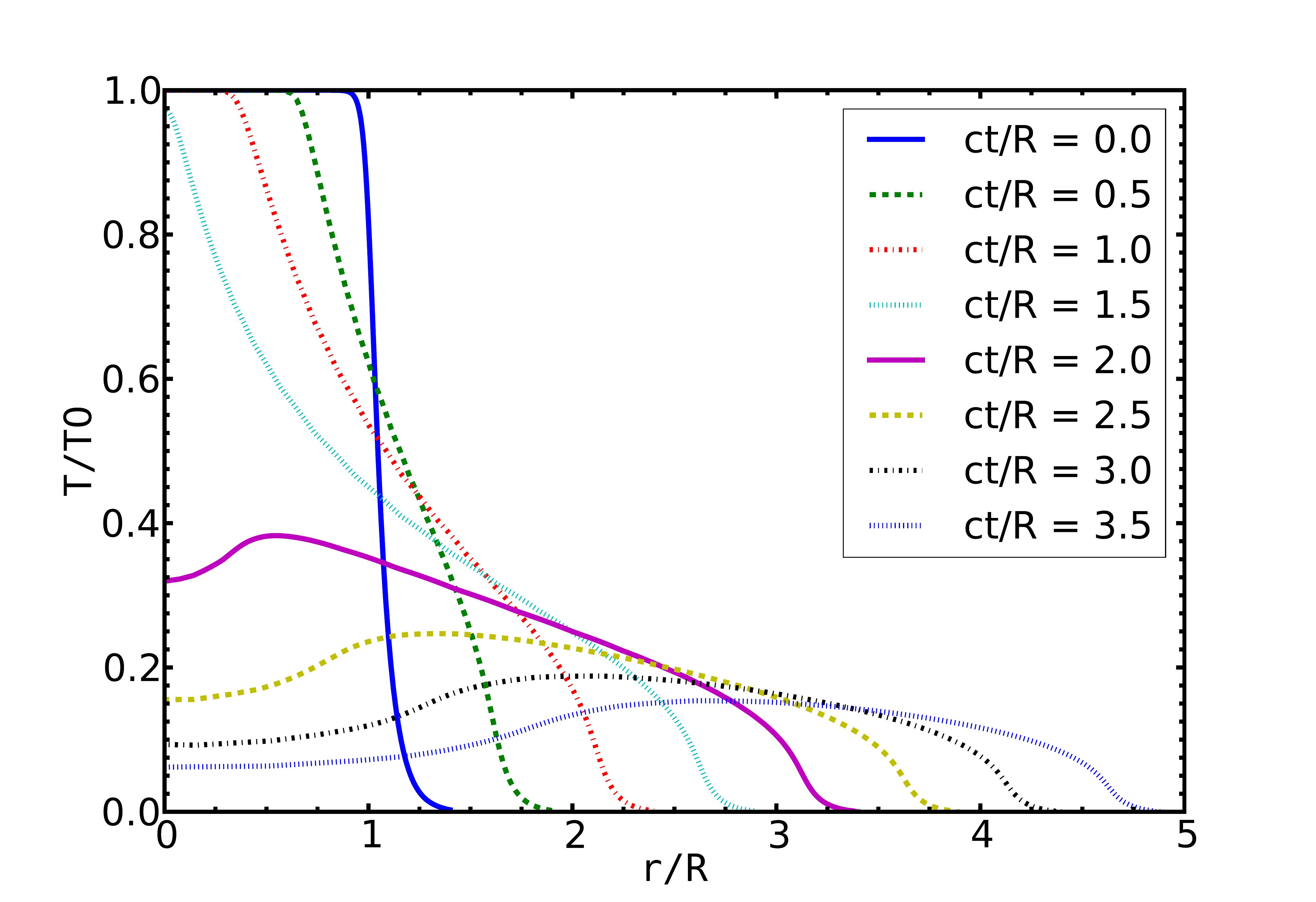}
\includegraphics[width=0.49\columnwidth]{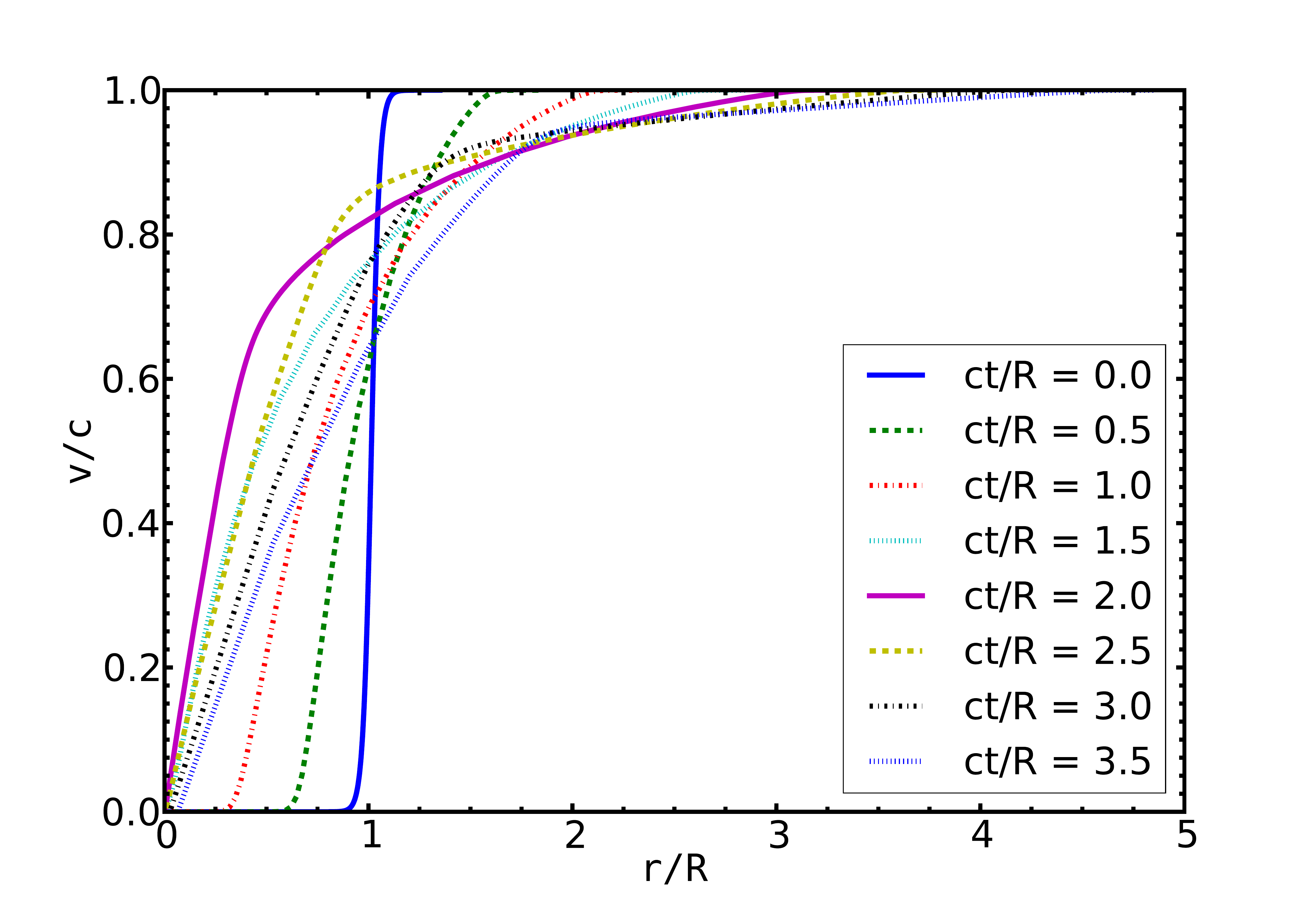}
\caption{\label{fig.2}
Temperature (left) and velocity (right) profiles at various time instants for the
spherically symmetric expansion.} 
\end{center}  
\end{figure} 

The solution for the given equation of state can be represented for scaled
quantities, $T/T_0$ and $r/R$, where $T_0$ and $R$ stand for the initial
temperature and radius, respectively, of the $e^- e^+ \gamma$ droplet. 
We assume initially homogenous droplets,
i.e.\ $T(t = 0, r) = T_0$ for $r < R$ and $T(t = 0, r) = 0$ for $r > R$.
The initial radial profile is $v(t = 0, r) = 0$ for $r < R$.
Initially, a rarefaction wave travels towards the center with the velocity of
sound $v_s = c / \sqrt{3}$
(corresponding to a Riemann type solution) and becomes reflected then,
causing the non-trivial radial temperature profile with somewhat lower
temperature in the center.
We allow the temperature evolution
until $\langle T \rangle \sim 0.1 T_0$. For late times, the temperature
drops further, but our assumed equation of state 
as well the photon rate below based on the perturbative estimates 
do not longer apply.
These late times until disassembly are not important for the
hard photon emission off the expanding droplet.
The scaled time to cool from $T_0$ to $\langle T \rangle \sim 0.1 T_0$
is about $ct = 3.5 R$.
That means, for instance, a droplet of initial radius $R = 2$ nm and $T_0 = 10$ MeV
with energy content 13 kJ expands within the relaxation time
for pairs and photon chemical equilibration,
0.025 fs, and cools to $\langle T \rangle \sim 1$ MeV.

The radial velocity profiles in Fig.~1 expose the rapid expansion of the cool
surface layers nearly at velocity of light.
In particular, due to the rapid expansion, 
the droplet cools faster that what was assumed in the
schematic model in \cite{Munshi_bk}. This is reflected in the photon
emission considered in the next section. 

\section{Photon emission\label{sec3}}

Having at our disposal the realistic temperature and velocity profiles,
we can now proceed to calculate the photon emission. It should be emphasized
that the leakage of energy via photon emission is not properly accounted
for in (\ref{eq.1}, \ref{eq.2})
(cf.~\cite{HW} for hydrodynamic schemes with energy leakage).
It turns out however that the emitted energy during the considered evolution
is small in comparison with the internal and kinetic energies so that
a consistent hydrodynamic treatment with radiation transport analog
to stellar core collapse simulations is not needed.

We employ the emission rate from thermo-field theory \cite{Moore}
(cf.\ figure 1 in \cite{Munshi_bk})
\begin{eqnarray} \label{eq.3}
\omega \frac{dN}{d^4x d^3 k} &=& 
\frac{2\alpha m^2_\infty f_{+1}(\omega)}{(2\pi)^3} 
\left [\ln \left (\frac{T}{m_\infty}\right )+\frac{1}{2} \ln \left ( 
\frac{2 \omega}{T} \right )
\right .  \nonumber  \\
&+&\left . C_{22}\left(\frac{\omega}{T}\right) \ 
+  \ C_{\rm{b}}\left(\frac{\omega}{T}\right) 
+  C_{\rm{a}}\left ( \frac{\omega}{T}\right) \right ],
\label{local}
\end{eqnarray}
where $f_{+1}(\omega) = (\exp (\omega/T)+1)^{-1}$ 
denotes the Fermi distribution function, 
and the dynamically generated asymptotic mass of the electron is 
$ m^2_\infty = 2 m^2_{\rm{th}}$.
The thermal mass squared is determined by $m^2_{\rm{th}}=e^2T^2/8$ 
with $e^2=4\pi\alpha$ and the $C$ coefficients are given explicitly 
in \cite{Munshi_bk}.
$\hat E_\gamma = u^\mu k^\nu g_{\mu \nu}$ 
takes properly into account the red/blue shifts
via the scalar product
of the medium's four-velocity $u^\mu(t, \vec x)$ and the photon's four-momentum 
$k^\nu = (\omega, \vec k)$
The rate includes $2 \leftrightarrow 2$ Compton and annihilation processes
(first three terms), bremsstrahlung processes $(C_b)$ and off-shell pair annihilations
processes ($C_a$). 

The space-time integrated emission spectrum includes only such photons
with a mean free path $\lambda_\gamma$ being larger larger than the 
distance to the cool droplet surface in a given direction.
The mean free path is locally determined by $\lambda_\gamma = \Gamma_\gamma^{-1}$
with the damping rate
$\Gamma_\gamma(\omega) = \frac{(2\pi)^3}{4} \ e^{\omega / T} \ \frac{dN}{d^4xd^3k}$
which depends in turn on the photon energy.

\begin{figure}[!htbp]
\begin{center}
\includegraphics[width=0.75\columnwidth]{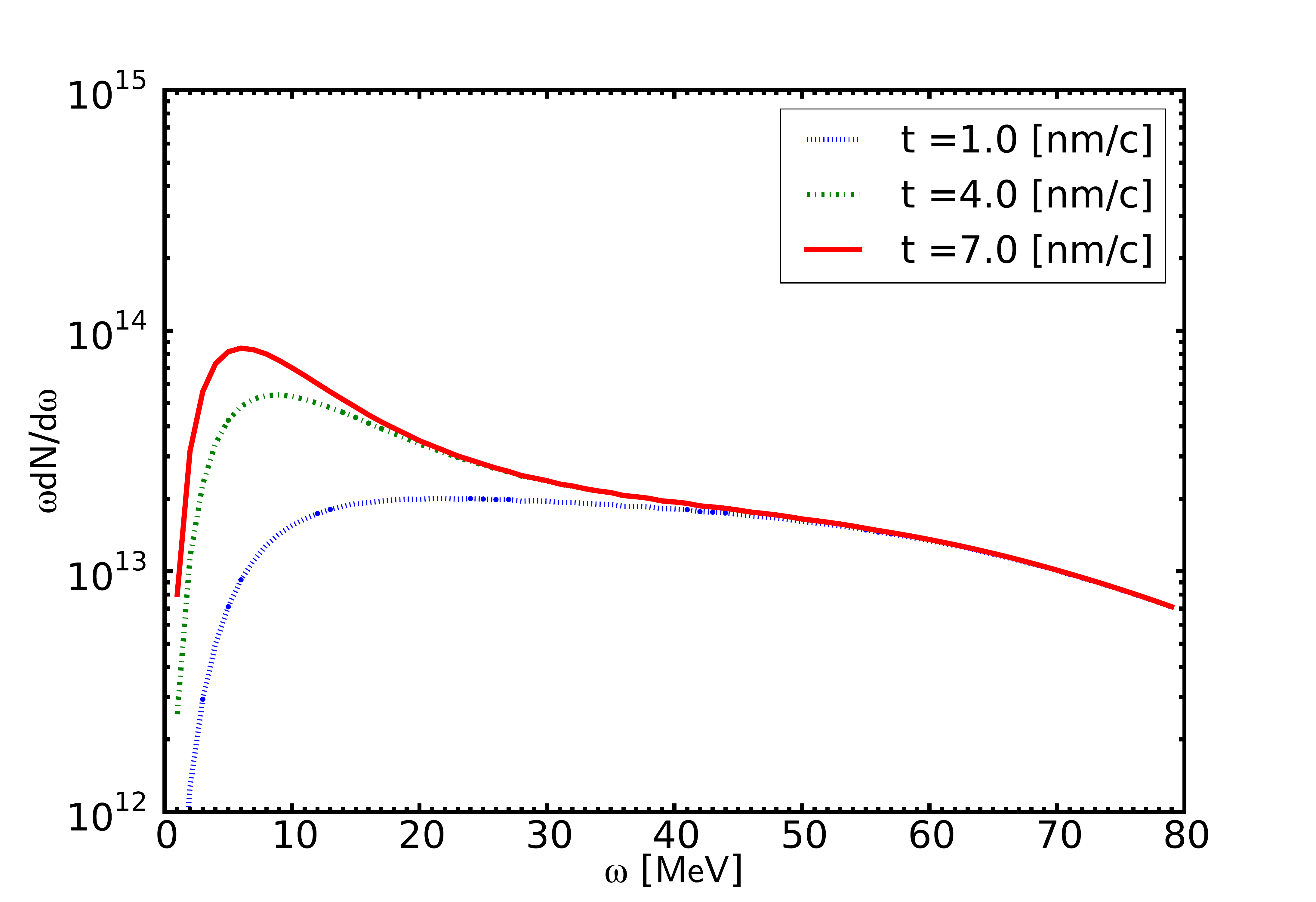}
\caption{\label{fig.3}
Energy-weighted spectrum of emitted photons as a function of the energy
for various times for $T > 1$ MeV.}
\end{center}  
\end{figure}

The resulting energy-weighted photon spectrum is exhibited in Fig.~\ref{fig.3}.
Similar to the schematic model \cite{Munshi_bk} it extends beyond 50 MeV with
notable strength. In contrast to \cite{Munshi_bk}, however, the maximum of the
spectrum is an order of magnitude smaller. The origin of this difference may be
traced back to the improved description of the droplet dynamics
which makes the droplet cooling considerably faster.
The hard tail of the photon spectrum is emitted at early times and does not change 
at later times. In contrast, the softer part with $\omega \sim 10$ MeV
gets significant contributions at later times. Comparing the spectra
at $t = 4$ nm/c and $t = 7$ nm/c, one concludes that the very
late radiation contribution for $T < 1$ MeV will contribute to photon
energies $\omega < 5$ MeV.    

\section{Broadening of the 511 keV Line\label{sec4}}

The photon emission rate (\ref{eq.3}) applies for $T > m$, i.e. a hot
plasma. In a warm plasma with $T \sim m$ or $T < m$, one expects
a significant contribution from the annihilation process
$e^- e^+ \to \gamma \gamma$. The corresponding inclusive photon rate 
for this process is obtained by the kinetic theory estimate 
\begin{eqnarray}
\omega \frac{dR}{d^4x \, d^3 k} &=&
\frac{{\cal N}}{(2\pi)^7 16 \omega}
\int_{4 m^2}^\infty \frac{ds}{s}
\int_{t_{min}}^{t_{max}} dt \,
\vert {\cal M} (s,t) \vert^2
\int_{\nu_{min}}^\infty d\nu \, (1 + f_{-1}(\nu-\omega)) \nonumber \\
&\times&  \int_{E_{min}}^{E_{max}} dE
\frac{f_{+1}(E) \, f_{+1} (\nu - E)}{\sqrt{(E_{max} - E)(E - E_{min})}}
\label{mass_rate}
\end{eqnarray}
with ${\cal N} = 4$ and
\begin{eqnarray}
t_{{min}\atop{max}} &=& m^2 - \frac12 s \left( 1 \pm \sqrt{1 - \frac{4 m^2}{s}} \right), \\
\nu_{min} &=& \omega + \frac{s}{4 \omega}, \quad t' = t -m^2,\\
E_{{min}\atop{max}} &=& \frac{-B \pm \sqrt{B^2 - 4 AC}}{2A},\\
A &=& - s^2, \quad B = 2s (\omega s + 2 \omega t' - \nu t'),\\
C &=& s t' (s+t') - \omega^2 s (s- 4 m^2) \nonumber\\
&-& 2\omega \nu s t' - \nu^2 t'^2
- 4 \omega \nu m^2 s + s^2 m^2
\end{eqnarray}
along with the additional kinematical restriction
$-4s^2 t' (s+t') \left(1+s m^2/((s+t')t')\right) > 0$. 
${\cal M}(s,t)$ is the lowest-order invariant matrix element for the
process $e^- e^+ \to \gamma \gamma$ (cf.~\cite{Rafelski2}),
\begin{equation}
\vert {\cal M} (s,t) \vert^2 = 32 \pi^2 \alpha^2 \left(
\frac{m^2 - u}{m^2 -s} +
\frac{m^2 - t}{m^2 - u} +
4 \frac{m^2}{m^2 -t} +
4 \frac{m^2}{m^2 -t} -
4 \left[ \frac{m^2}{m^2 -t} + \frac{m^2}{m^2 -u} \right]^2 \right)
\end{equation}
with $u = 2 m^2 - s -t$,
and the distribution functions are
$f_{\cal S}(y) = (\exp\{y/T\} + {\cal S})^{-1}$.  
For $T \gg m$, corresponding to $m\to 0$, one recovers the form presented in 
\cite{gamma_rate}; the rate scales $\propto T^2$. 
In the opposite region $\omega \gg T$, where the Boltzmann approximation
for the $f_{+1}$ terms is justified \cite{gamma_rate}, one finds
\begin{eqnarray}
\label{Boltzmann}
\omega \frac{dR}{d^4x \, d^3 k} &=&
\frac{{\cal N}}{2 (2\pi)^5} \frac{T}{\omega} {\rm e}^{-\omega/T} \nonumber \\
& \times &\int_{4 m^2}^\infty \frac{ds}{s} \, (s-4m^2)^2 \, \sigma (s) \, 
\ln (1-{\rm e}^{-s/(4 \omega T)} )^{-1},
\end{eqnarray}
where the total cross section $\sigma (s)$ follows from
integrating $d \sigma / dt =  \vert {\cal M} \vert^2 / (16 \pi (s - 4 m^2)^2)$
within $t_{min} - t_{max}$.
Due to the interplay of the factors ${\rm e}^{-\omega/T}$ and
$\ln (1-{\rm e}^{-s/(4 \omega T)} )^{-1}$ in (\ref{Boltzmann})
the rate has a sharp maximum
at $\omega = m$ for $T \ll m$. This is the 511 keV annihilation line,
at very low temperatures,
which broadens rapidly with increasing temperature. 
The high-frequency tail, $\omega \gg max(T,m)$, behaves as 
$\omega dR/d^3k \propto {\rm e}^{- \omega / T}$.
The full width at half maximum is 
$\Delta \omega = \sqrt{-4 m T \ln \frac12}$, for $T \ll m$,
with maximum $\propto {\rm e }^{-2m/T}$.

\begin{figure}[!htbp]
\begin{center}
\includegraphics[width=0.75\columnwidth]{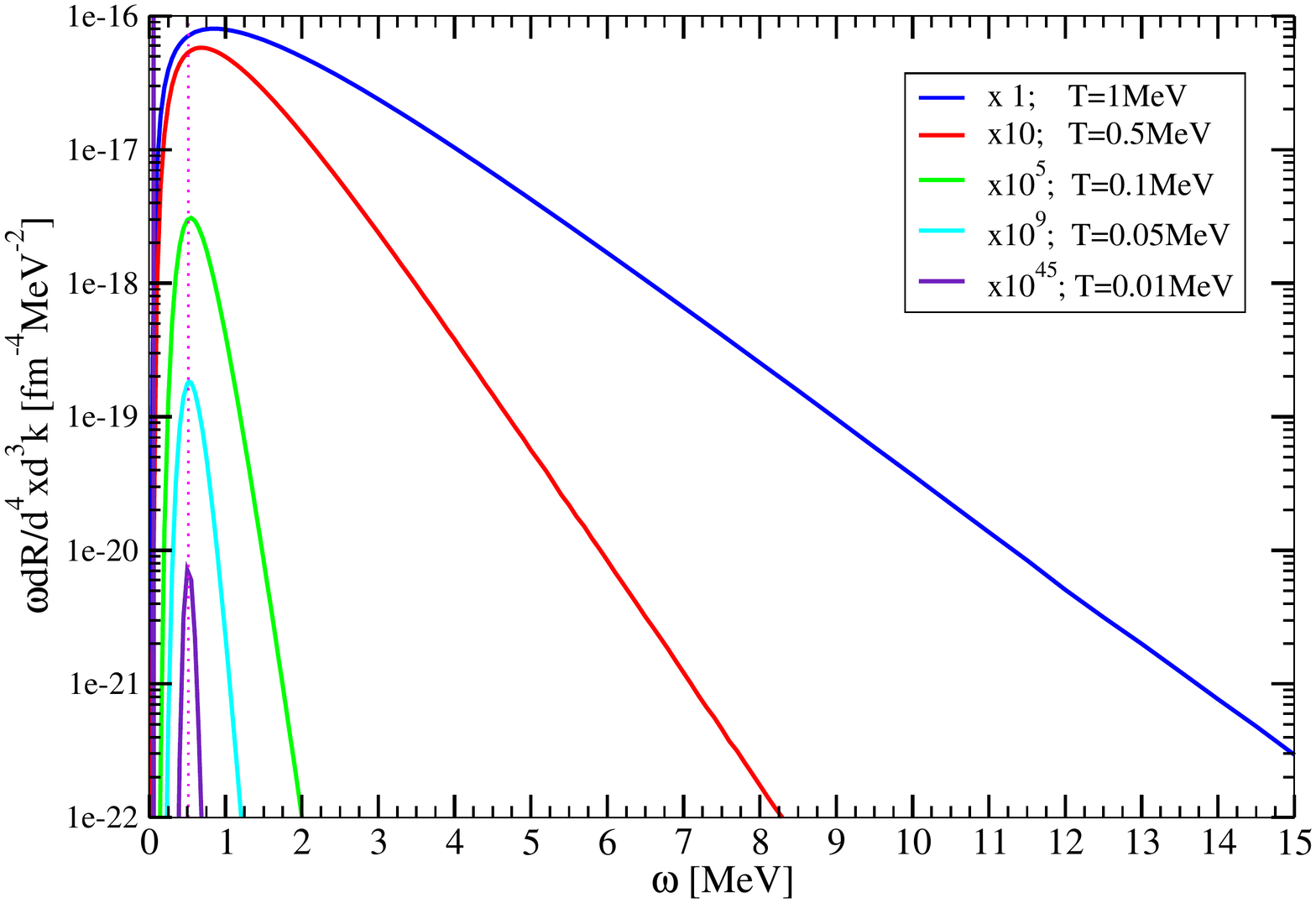}
\caption{\label{fig.A3}
Photon emission rate from $e^- e^+$ pair annihilation 
as a function of photon energy for various values of the temperature.
Note the scale factors in the legend. 
The vertical dashed line indicates the position of 511 keV line.} 
\end{center}  
\end{figure}

The rate (\ref{mass_rate}) is displayed in Fig.~\ref{fig.A3} for a sequence of temperatures.
With increasing temperature the relative kinetic energy of the annihilating
electron-positron pairs increases and, consequently, the photon energy
increases in the medium's rest frame and becomes larger than $m$. This
broadening of the 511 keV annihilation line, once isolated from other 
processes with continuum radiation such as the Compton and the various
bremsstrahlungs, may serve as a measure 
of the achieved temperature in a warm $e^- e^+ \gamma$ plasma.
At $T > 0.1 m$, the broadened annihilation line changes gradually into a smooth
continuum with typical exponential tail. For $T > m$, the maximum of the
emission rate is shifted to $\omega > m$. The annihilation is then part
of the complete rate (\ref{eq.3}).

The presented annihilation rates apply also to the transiently created
$e^- e^+$ pairs by the dynamical Schwinger process, as envisaged in \cite{Blaschke},
when approximating the distribution functions by thermal ones. 

\section{Summary and Discussion\label{sec5}}

In summary we consider the expansion dynamics of hot
plasma droplets which can be created with next-generation
ultra-intense laser beams. We present here the dynamics of charge-symmetric and
equilibrium $e^- e^+ \gamma$ plasma droplets by means of relativistic hydrodynamics.
Our results support the estimate in \cite{Munshi_bk} pointing to gamma flashes
emitted off the rapidly exploding droplets. 

Clearly, our considered scenario is still idealized. The surplus of electrons
over positrons needs to be considered in future analysis. This can be
accommodated by a finite chemical potential steering the asymmetry of electrons 
and positrons. Furthermore, at given temperature, implying kinetic equilibrium,
the positrons may not be chemically equilibrated. This requires an additional
dynamical variable, e.g.\ a positron fugacity supplemented by appropriate
rate equations. Also, ion impurities may have an impact on the plasma dynamics.

As indicated by the largely differing expansion time scales for spherical and plane
expansion patterns, the flow symmetry may play an important role. To clarify
that issue, more simulations 
are needed along with realistic initial conditions. The lessons learnt
from the investigation of the quark-gluon plasma \cite{QM}, however, indicate
that it is useful to develop parallel with experiments the theoretical description
after first initializing considerations. 

Finally, we mention the broadening of the 511 keV line as useful tool for
thermometry of small electron-positron-photon droplets produced in
the laboratory and in astrophysical sites.  

\acknowledgements
The authors thank T.E. Cowan and R. Sauerbrey for inspiring discussions.
MGM is thankful to FZ Dresden-Rossendorf for support during the 
course of this work.

\appendix
\section{Planar Slab Geometry\label{appA}}

\begin{figure}[!htbp]
\begin{center}
\includegraphics[width=0.49\columnwidth]{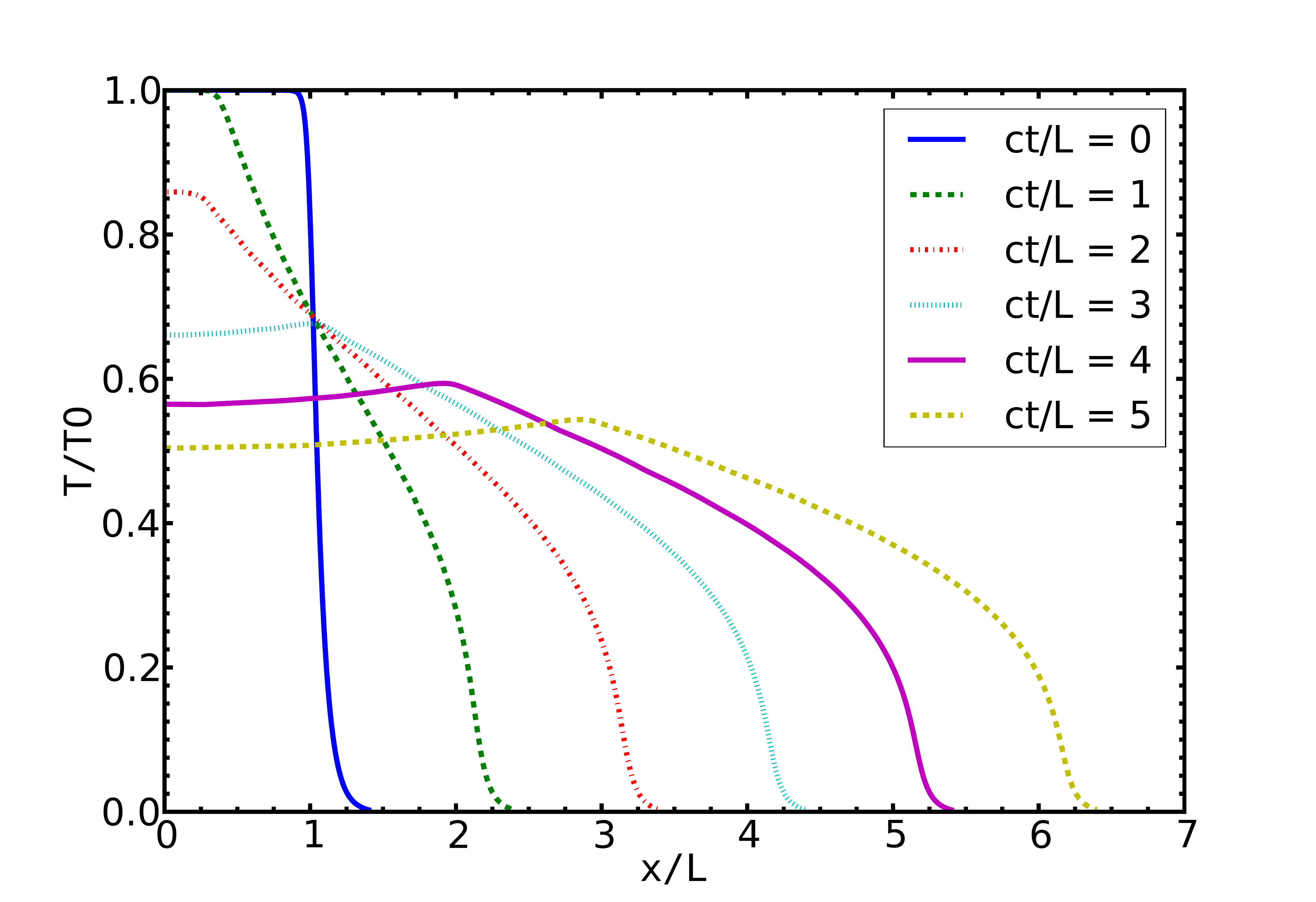}
\includegraphics[width=0.49\columnwidth]{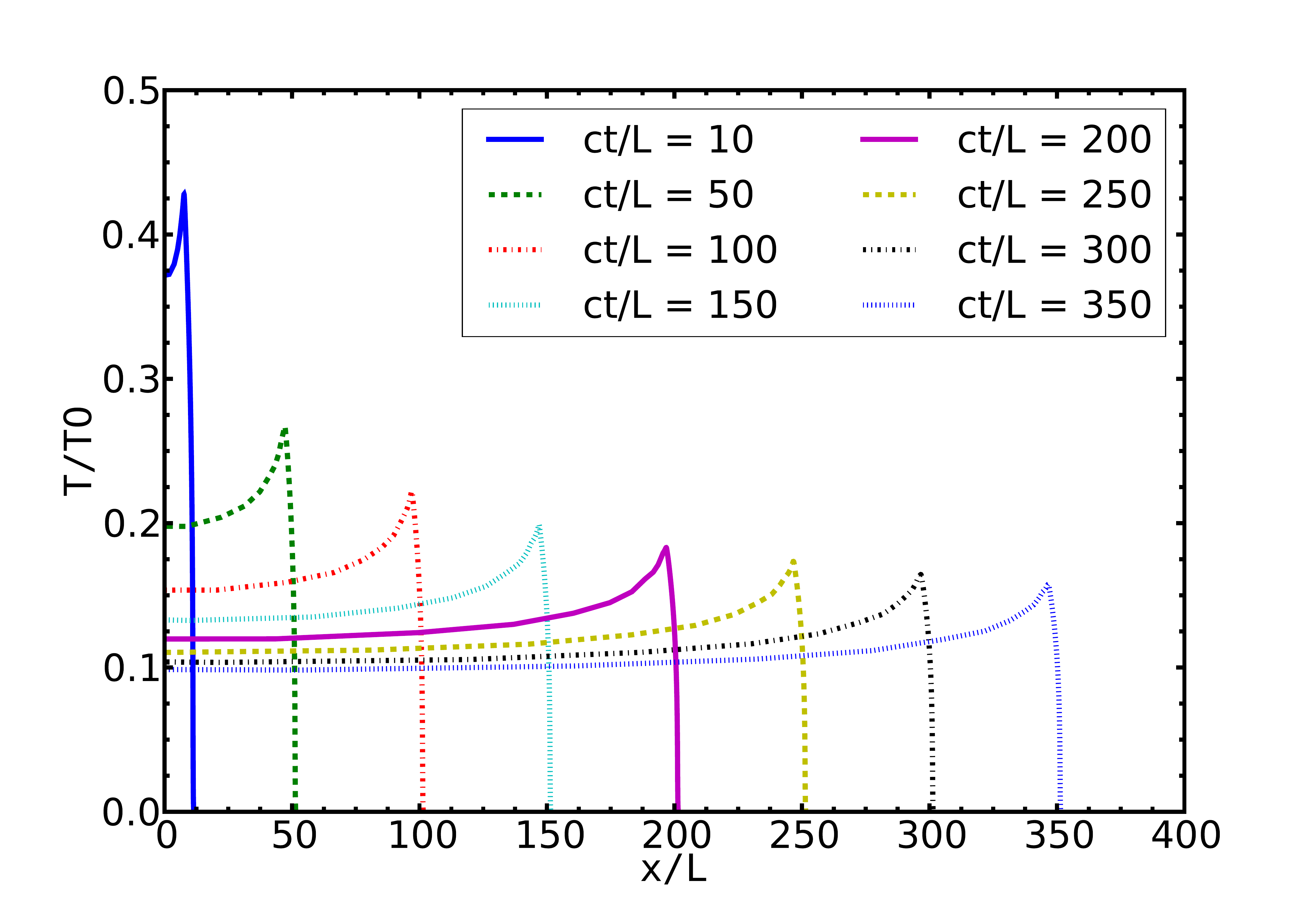}
\caption{\label{fig.A1}
Temperature profiles at various time instants for a planar
one-dimensional expansion. Left (right): short (long) times.} 
\end{center}  
\end{figure}

\begin{figure}[!htbp]
\begin{center}
\includegraphics[width=0.49\columnwidth]{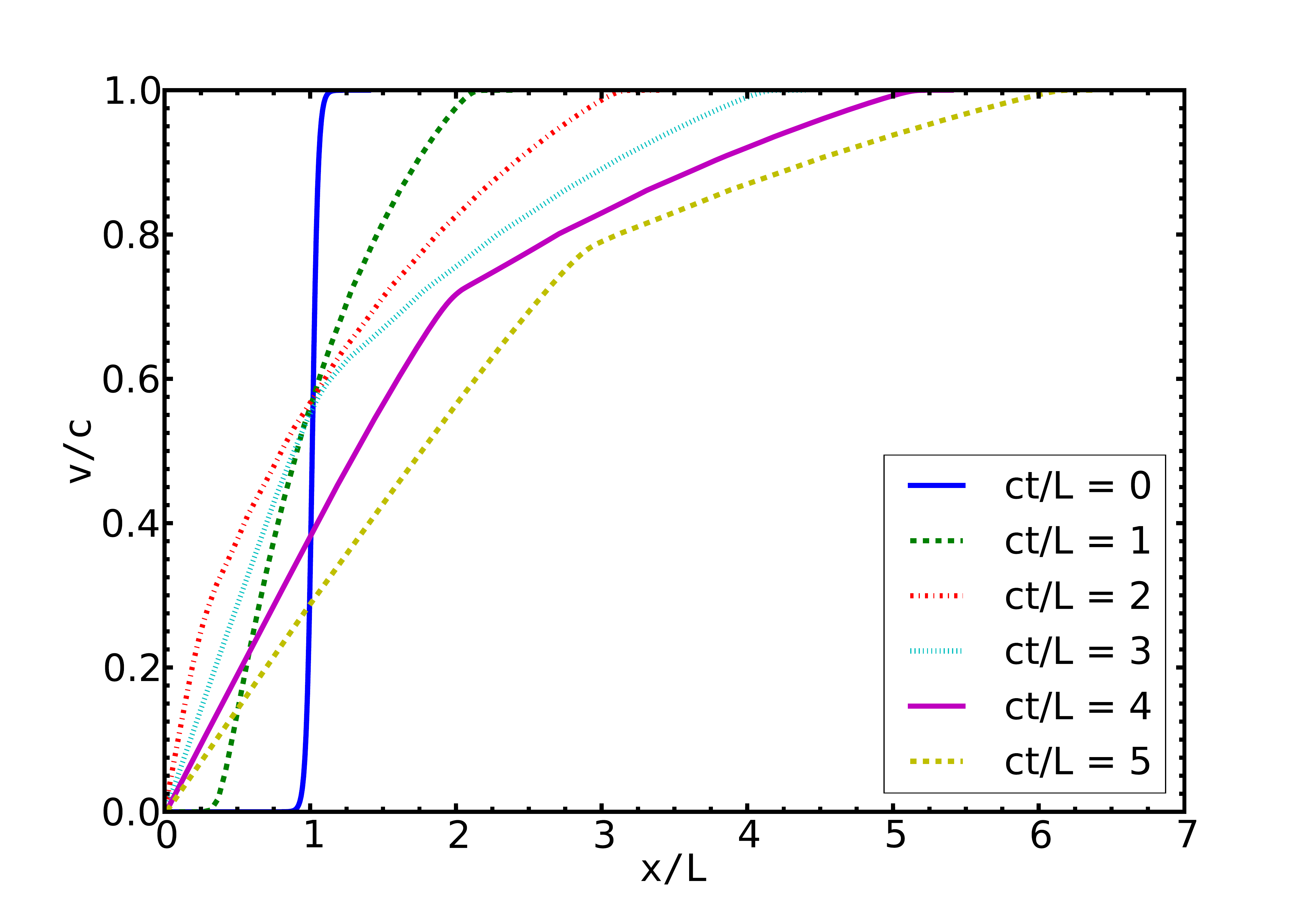}
\includegraphics[width=0.49\columnwidth]{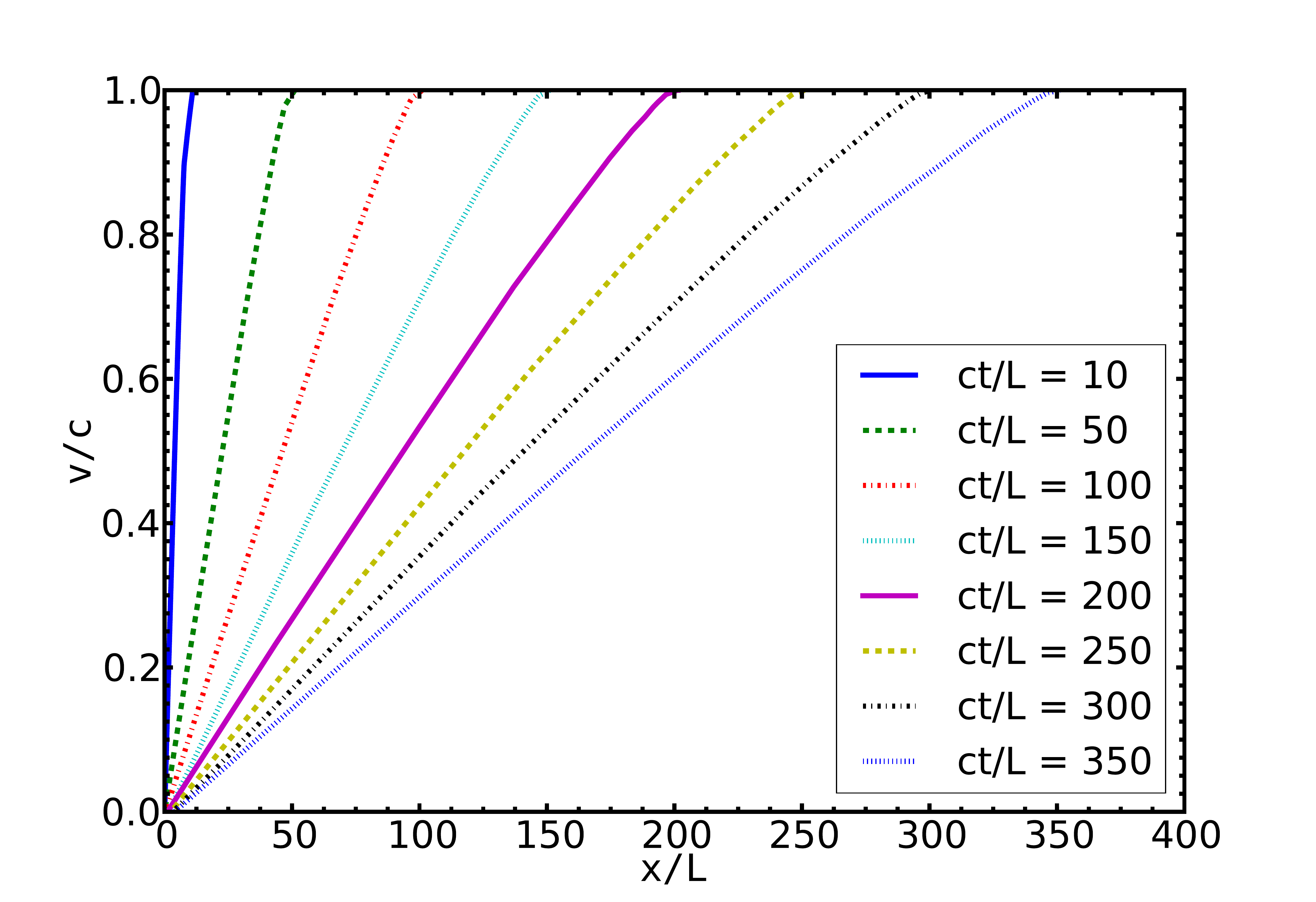}
\caption{\label{fig.A2}
Velocity profiles at various time instants for a planar
one-dimensional expansion. Left (right): short (long) times.} 
\end{center}  
\end{figure}

A one-dimensional planar expansion (cf.~\cite{Baym}) needs a much longer time to cool
the system down to $T_0$ to $\langle T \rangle \sim 0.1 T_0$.
This is exhibited in Figs.~\ref{fig.A1} and \ref{fig.A2}.

To understand qualitatively this huge difference to the spherically
symmetric expansion, one may mention that for the latter one an increase
of the radius by a factor of $10^{4/3}$ is needed to cool the medium from
$T_0$ to $0.1 T_0$ when neglecting the conversion of internal energy into kinetic
energy and emission. For planar geometry, in contrast, the thickness of a slab
must increase by a factor $10^4$. This estimates explain the much longer
time scales to cool the medium to a tenth of the initial temperature.
Of course, the transverse expansion of a finite slab of matter
will become important and requires more involved multi-parameter
solutions of the hydrodynamical expansion.        

\end{document}